\documentclass{article} %% comment this out for double column
\usepackage{upgreek}
\usepackage{mathtools}
\usepackage{rotating}
\usepackage{amsmath}
\usepackage{amssymb}
\usepackage{bbm}
\usepackage{subcaption}
\usepackage{enumitem}

\usepackage[authoryear,longnamesfirst]{natbib}

\def\bSig\mathbf{\Sigma}

\newcommand{\un}[1]{\boldsymbol{#1}}
\newcommand{\pbi}{\begin{itemize}}
	\newcommand{\pei}{\end{itemize}}

\newcommand{\pbc}{\begin{center}}
	\newcommand{\pec}{\end{center}}
\newcommand{\pbe}{\begin{eqnarray*}}
	\newcommand{\pee}{\end{eqnarray*}}
\newcommand{\pben}{\begin{eqnarray}}
	\newcommand{\peen}{\end{eqnarray}}
\newcommand\argmax[1]{\underset{#1}{\mathrm{argmax} \text{ }}}

\usepackage{tikz}
\usetikzlibrary{arrows}
\begin{document}

%\let\WriteBookmarks\relax
%\def\floatpagepagefraction{1}
%\def\textpagefraction{.001}
%\shorttitle{Multivariate spatial conditional extremes}
%\shortauthors{R. Shooter et~al.}
%
\title{Multivariate spatial conditional extremes for extreme ocean environments}
\author{Rob Shooter, Emma Ross, Agustinus Ribal, Ian R. Young, Philip Jonathan}
\date{Version accepted for publication in Ocean Engineering, January 2022}
\maketitle

\begin{abstract}
	The joint extremal spatial dependence of wind speed and significant wave height in the North East Atlantic is quantified using Metop satellite scatterometer and hindcast observations for the period 2007-2018, and a multivariate spatial conditional extremes (MSCE) model, ultimately motivated by the work of Heffernan and Tawn (2004). The analysis involves (a) registering individual satellite swaths and corresponding hindcast data onto a template transect (running approximately north-east to south-west, between the British Isles and Iceland), (b) non-stationary directional-seasonal marginal extreme value analysis at a set of registration locations on the transect, (c) transformation from physical to standard Laplace scale using the fitted marginal model, (d) estimation of the MSCE model on the set of registration locations, and assessment of quality of model fit. A joint model is estimated for three spatial quantities: Metop wind speed, hindcast wind speed and hindcast significant wave height. Results suggest that, when conditioning on extreme Metop wind speed, extremal spatial dependence for all three quantities decays over approximately 600-800 km.
\end{abstract}

\section{Introduction}
Severe ocean events at a location often involve extreme values for more than one oceanographic variable, leading to interest in the development and tailoring of statistical methods for multivariate and conditional extremes in a met-ocean context; for example, extreme loads on an offshore structure often arise from a combination of wind, wave and current. Further, simultaneous occurrences of extreme ocean events at different locations often represent higher risk than extreme events at one location; for example, extreme storm events often impact multiple structures simultaneously, requiring integrated planning of activities such as unmanning. Storms load coastal defences over a spatial neighbourhood, rather than at a single location. Design and maintenance of a wind farms requires characterisation of the joint spatial field for metocean variables over a spatial neighbourhood. This leads to interest in the development of statistical models for the joint characteristics of multiple spatial fields of variables such as wind speed and significant wave height. The objective of this article is to develop a straightforward model for Multivariate Spatial Conditional Extremes (MSCE), and demonstrate its usefulness in met-ocean application. Specifically, we seek a model for the joint behaviour of multiple met-ocean variables in space, given an occurrence of an extreme value of one of the variables at some location. The idea underpinning the methodology developed is the conditional extremes model of \cite{HffTwn04}. As discussed in Section~\ref{Sct:Mth}, the MSCE model can be seen as the latest in a sequence of extensions of the conditional extremes approach to address specific applications' requirements. The main use of the MSCE model is to provide a principled approach to characterising the joint structure of spatial fields, to be applied in any situation where environmental hazard results from extreme values of at least one metocean variable.

Like its conditional extremes predecessor, inference using MSCE amounts essentially to fitting a non-linear regression model for observations of random variables on a standard Laplace marginal scale. The model admits different types of extremal dependence (including asymptotic dependence and asymptotic independence, \citealt{ClsHffTwn99}) and is computationally rather straightforward to estimate. 

Other statistical approaches to spatial extremes are motivated by the theory of max-stable processes (MSPs; see e.g., \citealt{BrnRsn77}, \citealt{Smt90}, \citealt{Sch02}, \citealt{DvsPdnRbt12}, \citealt{Rbt13} and the recent review of \citealt{HsrWds20}). However, typical MSP models require an assumption that the extremal spatial dependence takes a particular form. Some models in principle are able to permit different classes of extremal dependence (e.g., \citealt{WdsEA17}) but can be fairly restrictive or computationally unwieldy. 

Here, we are interested not only in quantifying the extremal spatial dependence for quantities such as wind speed and significant wave height, but also the cross-dependence between different quantities. In this context, the hierarchical max-stable spatial model of \cite{RchShb12} has been extended (\citealt{Vtt17}, \cite{RchShb18}, \citealt{VttHsrGnt19}) to the multivariate case. Further, \cite{GntPdnSng15} introduced multivariate max-stable spatial processes. These approaches certainly have their merits, but also have at least some of the limitations outlined above pertaining to MSP models.

Estimation using the MSCE model for a sample of spatial data for a number of different quantities, first requires that the data are presented on a common marginal standard Laplace scale for all quantities at all locations. This transformation is achieved by estimating non-stationary directional-seasonal extreme value models for each quantity at each location (e.g. \citealt{ZnnEA19a}), and then application of the probability integral transform.

The layout of the article is as follows. Section~\ref{Sct:MtvApp} provides a description of the motivating application to wind speed and significant wave height in the North East Atlantic, satellite and hindcast data sources, and data pre-processing steps. Section~\ref{Sct:Mth} then presents the MSCE methodology. Results of applying the MSCE model in the North East Atlantic are given in Section~\ref{Sct:Rsl}, and Section~\ref{Sct:Dsc} provides a discussion. Supporting plots for the exploratory data analysis are given in the appendix.

\section{Motivating application} \label{Sct:MtvApp}
We seek to characterise the joint spatial conditional structure of extreme values of wind speed and significant wave height in the North East Atlantic between the British Isles and Iceland, subject to systematic directional and seasonal variability. In this section, we describe data sources, and data pre-processing necessary prior to MSCE inference. Two sources of wind data are available, namely satellite-observed wind speed \texttt{StlWnd} and direction, and hindcast model wind speed \texttt{HndWnd} and direction. Hindcast significant wave height \texttt{HndWav} and wave direction are also used. These data sources are combined to yield a set of observations of \texttt{StlWnd}, \texttt{HndWnd} and \texttt{HndWav} and corresponding directional and seasonal data, on a transect of equally-spaced registration locations lying between approximately $2^\circ$W, $67^\circ$N and $22^\circ$W, $55^\circ$N. 

The objective of the analysis is to evaluate our ability to quantify the joint structure of spatial fields corresponding to significant wave height and wind speed, based on hindcast and satellite data, conditional on the occurrence of an extreme wind speed or significant wave height event. For the North East Atlantic, one use case for the model would be joint assessment of extreme loads on multiple offshore facilities, requiring joint characterisation of wind and wave fields, critical e.g. in the planning and execution of operations such as unmanning (\citealt{TowEA21}).

The appeal of direct earth observation by satellite is that it may provide, in the medium-to-long term, a rich source of data to inform ocean-related human activities in real time. At present, satellite data alone is not sufficient for design purposes of course because of poor temporal coverage and length of historical time-series for a location; hindcast data is clearly a more useful data source at present. However, the quality of hindcast output for extreme values of waves and wind at a location is also uncertain, usually requiring calibration to local measurements, especially to accommodate directional and other covariate effects. Similarly, satellite output currently requires calibration. For all these reasons, it is important to quantify how inferences for extremes from satellite observations compare with those from hindcasts, and to consider inferences for extremes which combine hindcast and satellite input.

\subsection*{Metop and hindcast data sources}

Metop-A, B and C are polar-orbiting meteorological satellites, forming the EUMETSAT Polar System (EPS) series. Metop-A (launched 19 October 2006), Metop-B (launched 17 September 2012) and Metop-C (launched 7 November 2018) are in a low polar orbit, at an altitude of approximately 800 km over the Earth's surface. Metop uses the Advanced SCATterometer (ASCAT) to measure wind speed and direction over the oceans. ASCAT is a real aperture radar, operating at 5.255 GHz (C-band) and using vertically polarised antennas. It transmits long pulses of microwave energy with ``chirp'' linear frequency modulation towards the sea surface. Winds over the sea cause centimetre scale disturbances of the sea surface which modify radar back-scattering characteristics dependent on both wind speed and direction. Two sets of three antennas measure the characteristics of the back-scattered signal in two 500 km wide swaths to each side of the satellite ground track, make sequential observations of the back-scattering coefficient of each point of interest from three directions. The characteristics of the back-scattered signals allow estimation of surface wind speed and direction. Over the North East Atlantic, the daily pass of each Metop satellite is from the north-east to the south-west. All Metop satellite data for the current work were sourced from the Australian Ocean Data Network (AODN). These datasets have been calibrated and quality controlled as described by \cite{RblYng20} and \cite{RblYng20a}. At wind speeds above 30m/s scatterometers tend to saturate (\citealt{RblTmzYng21}). However, as shown by \cite{RblYng20}, Metop-A and Metop-B yield unbiased data for wind speeds up to a minimum of 25m/s. Note that only Metop-A and Metop-B data are considered here.

The NORA10 hindcast (NOrwegion ReAnalysis 10 km, \citealt{RstEA11}) is a combined high-resolution atmospheric downscaling and wave hindcast for the Norwegian Sea, the North Sea and the Barents Sea, based on the European Reanalysis dataset (ERA-40), outputting 3-hourly wave fields at a resolution of 10 km for the period 1957–2018.

\subsection*{Registered data}
\begin{figure}[ht]
	\centering
	\includegraphics[width=0.5\columnwidth]{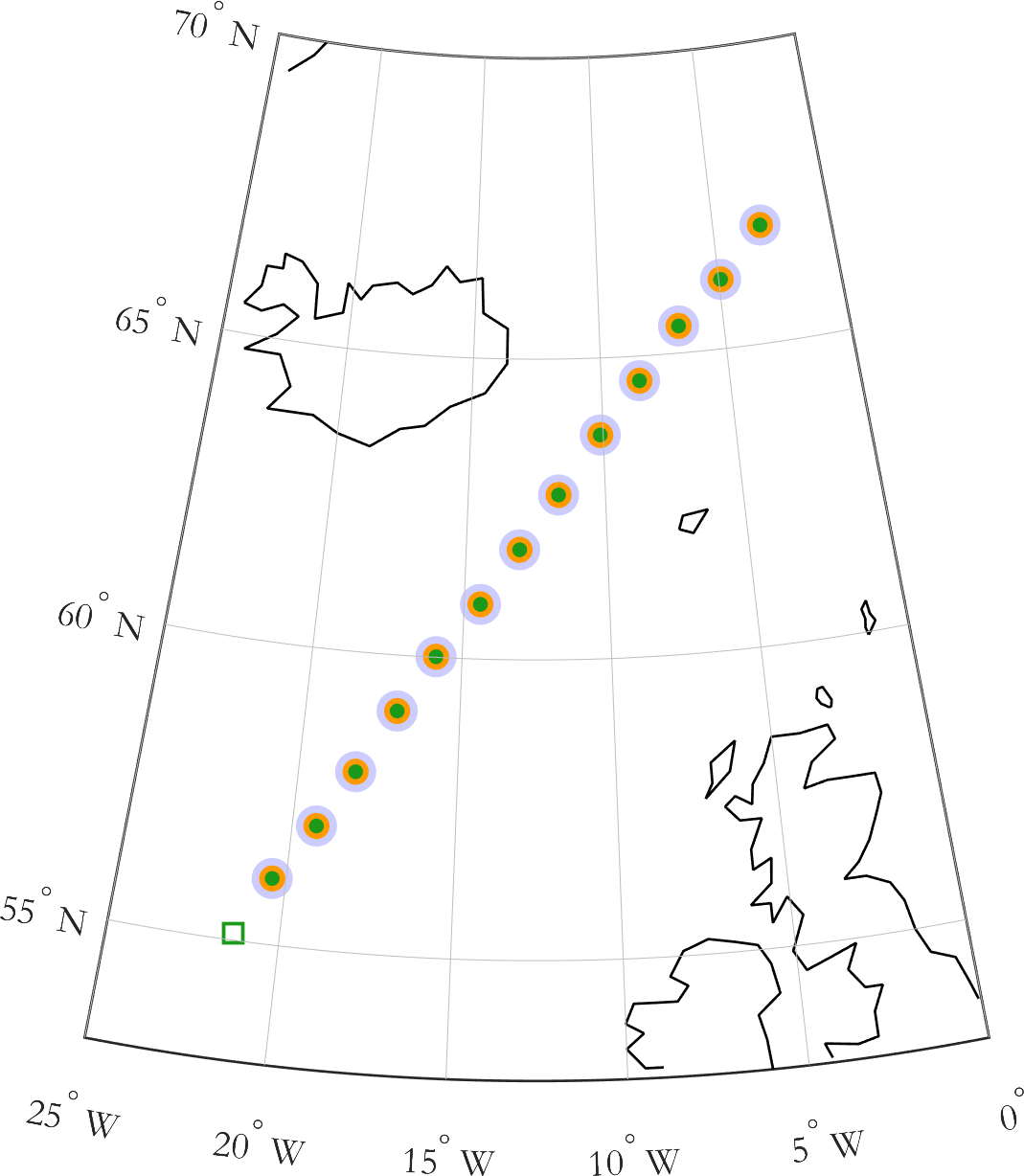}		
	\caption{Map of the registration locations. Squares indicate the conditioning location $r_j$, $j=0$, and discs indicate other (remote) locations $r_j$, $j=1,2,...,13$ with increasing distance from location $r_0$. Colour coding indicates that the conditioning quantity is \texttt{StlWnd} (green square, at location $r_0$), and that all of \texttt{StlWnd} (green), \texttt{HndWnd} (orange) and \texttt{HndWav} (blue) are used as conditioned quantities (discs) at remote locations. The Metop satellite trajectories are from north-east to south-west. (The green-orange-blue colour scheme appears as three shades of grey with decreasing intensity when viewed in black-and-white.)}
	\label{Fgr1}
\end{figure}

Figure~\ref{Fgr1} gives the locations of 14 equally-spaced registration locations of the registration transect selected, lying in the North East Atlantic between the British Isles and Iceland. The end points of the registration transect were chosen so that the transect covers the greatest number of satellite passes, thus maximising the size of the Metop sample for MSCE analysis. For each available satellite pass, we find the nearest point on the satellite transect to each of the registration locations, and allocate the value of wind speed and direction from the ``matched'' satellite location to the registration location. If the maximum ``matched'' distance (calculated using the spherical law of cosines) for any registration location corresponding to a specific satellite pass exceeds 50 km, the pass is not registered. Hindcast wind speed and direction, significant wave height and wave direction are registered similarly, based on matching spatially between the hindcast grid and the registration transect at the times of already-registered satellite passes. In this way, a total of 1532 joint observations of \texttt{StlWnd}, \texttt{HndWnd} and \texttt{HndWav} were isolated for analysis.

Referring to Figure~\ref{Fgr1}, the south-west location is taken as the conditioning location (square), and all other locations (discs) are used as remote locations. The colour-coding scheme used indicates that the conditioning quantity is \texttt{StlWnd} (green) at the conditioning location, but that all of \texttt{StlWnd} (green), \texttt{HndWnd} (orange) and \texttt{HndWav} (blue) are present in the MSCE model as conditioned variates at remote locations.

\begin{figure}[ht]
	\centering
	\includegraphics[width=0.8\columnwidth]{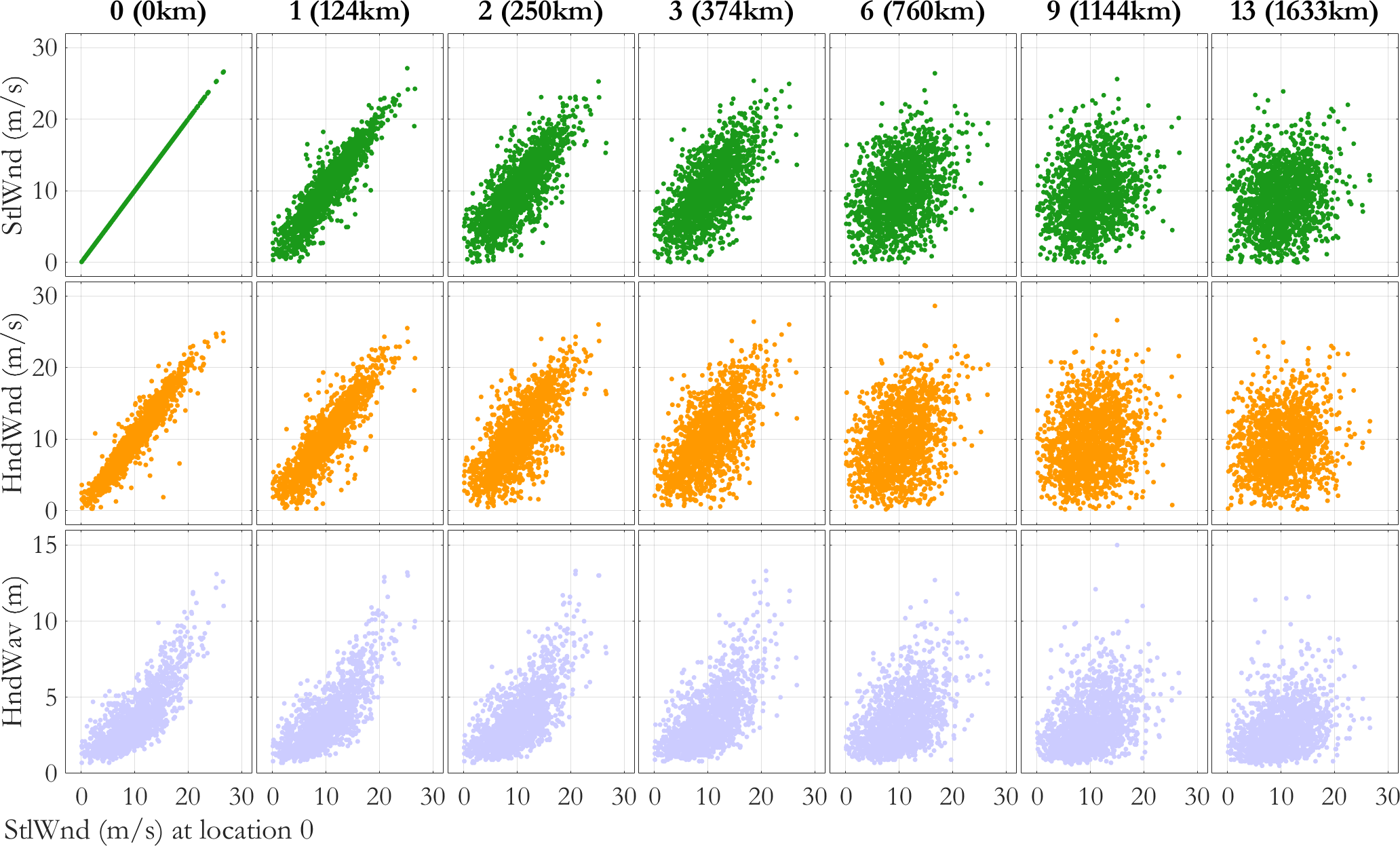}		
	\caption{Scatter plots of \texttt{StlWnd} (green), \texttt{HndWnd} (orange) and \texttt{HndWav} (blue; all on physical scale) at selected locations $r_j$, $j=0,1,2,3,6,9,13$ against \texttt{StlWnd} at the conditioning location $r_j$, $j=0$.}
	\label{Fgr2}
\end{figure}

 Figure~\ref{Fgr2} shows the dependence between each of these quantities at selected remote locations and the conditioning variate \texttt{StlWnd} at the conditioning location $r_j$, $j=0$. The dependence between \texttt{StlWnd} at a given remote location ($r_j$, $j=1,2,...,13$), and \texttt{StlWnd} at location $r_0$, is similar to that between \texttt{HndWnd} at location $r_j$ and \texttt{StlWnd} at location $r_0$. There is evidence for curvature in the relationship between \texttt{HndWav} at location $r_j$ and \texttt{StlWnd} at location $r_0$, reflecting the typical drag-type squared relationship between wind forcing and resulting significant wave height. There is also evidence that the dependence between \texttt{HndWav} at location $r_j$ and \texttt{StlWnd} at location $r_0$ decays more slowly with increasing distance (or increasing $r_j$) than that between wind speed at $r_j$ and \texttt{StlWnd} at location $r_0$. Again, this is indicative of greater spatial dependence for significant wave height in general compared to wind speed. Figures~\ref{FgrA1}-\ref{FgrA3} in the appendix provide supporting scatter plots illustrating the spatial dependence for each of \texttt{StlWnd}, \texttt{HndWnd} and \texttt{HndWav} fields separately.

\subsection*{Marginal transformation to standard Laplace scale}
\begin{figure}[ht]
	\centering
	\includegraphics[width=0.8\columnwidth]{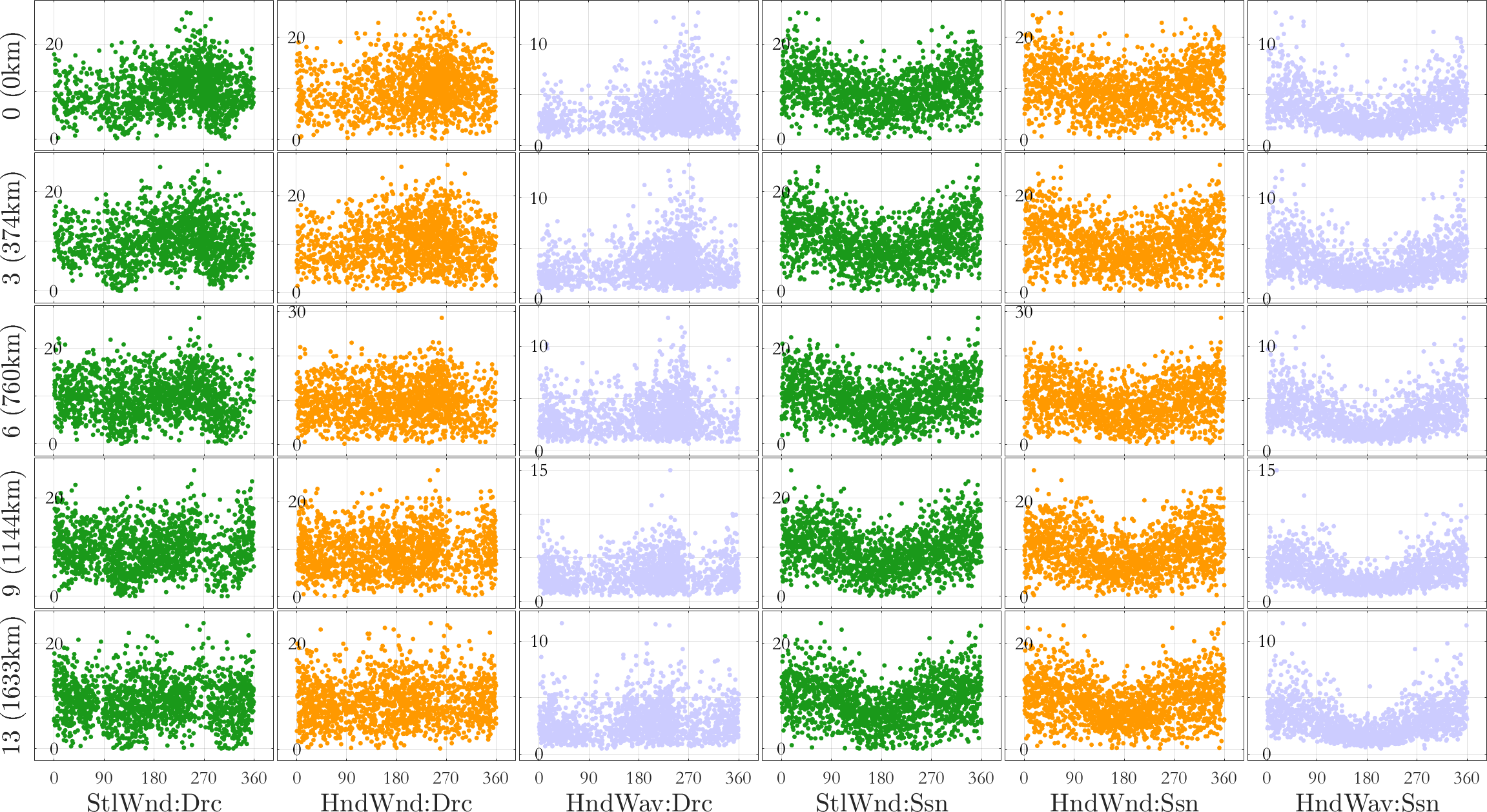}		
	\caption{Directional and seasonal covariate effects in registered data (on physical scale) for selected locations. The first three columns give directional variation for each of \texttt{StlWnd} (green), \texttt{HndWnd} (orange) and \texttt{HndWav} (blue) at locations $r_j$, $j=0, 3, 6, 9, 13$. The corresponding seasonal variation is illustrated in the fourth to sixth columns. Direction is defined as the direction from which fluid flows, measured clockwise from north. Season is defined as the day of the year (1,2,...,365 or 1,2,...,366 for leap years) mapped linearly on to the interval $(0,360]$.}
	\label{Fgr3}
\end{figure}

The registered data also show systematic variation with direction and season, as illustrated in Figure~\ref{Fgr3}. Seasonal effects on all three quantities are clear in columns 4-6 of the figure, for all locations considered. The directional effect is more interesting. For locations $r_0$, $r_3$ and $r_6$, the prevailing direction for wind and $H_S$ is from the west as might be expected in the north Atlantic. However, at locations $r_9$ and $r_{13}$, the directional dependence is more confused; here for significant wave height, there is evidence for large values emanating from the Norwegian Sea to the north or from the North Sea to the south.

Figure~\ref{Fgr3} makes clear that, to transform the registered data from physical to standard Laplace scale ready for MSCE inference requires fitting a non-stationary marginal model to the data for each of \texttt{StlWnd}, \texttt{HndWnd} and \texttt{HndWav} at each registration location, so that the effects of directional and seasonal variation are captured adequately. Non-stationary directional-seasonal marginal extreme value models are estimated using the PPC (penalised piecewise constant) extreme value methodology, described previously in \cite{RssEA17b}. Details of the analysis are withheld for brevity, in favour of the brief description here. The directional covariate domain is partitioned into eight octants centred on cardinal and semi-cardinal directions, and the seasonal covariate domain partitioned into ``summer'' and ``winter'' intervals centred on ``seasonal degrees'' $0^\circ$ and $180^\circ$. This yields a partition of the full directional-seasonal covariate domain into 16 directional-seasonal ``bins''. Then a piecewise constant generalised Pareto extreme value model for peaks over threshold is estimated simultaneously for all bins, such that the generalised Pareto shape parameter is constant everywhere on the covariate domain, and the generalised Pareto scale parameter is assumed constant within each bin. Further, the variation of the generalised Pareto scale parameter between bins is penalised. The extent of penalisation is regulated using cross-validation, to give the best ``out-of-sample'' predictive performance using the model. Moreover, a bootstrap scheme admitting a range of different plausible extreme value thresholds is incorporated, such that model uncertainty can be quantified reasonably. Software for the analysis is available at \cite{ECSADESGitHub}.

Models are estimated independently for each of \texttt{StlWnd}, \texttt{HndWnd} and \texttt{HndWav} at each registration location. Using simulation under the fitted model, the 100-year maximum value of \texttt{StlWnd} is around 35 ms$^{-1}$, with 95\% uncertainty interval of approximately (28,45) ms$^{-1}$ at each of the registration locations. The corresponding values for \texttt{HndWnd} are similar. For \texttt{HndWav}, there is evidence that the 100-year maximum value reduces from around 22 m,(16,30) m at south-western locations (corresponding to small $r_j$) to around 17 m,(13,26) m at north-eastern locations. These values are in general agreement with expectations and previous estimates at these locations.

Using the fitted models with bootstrap median parameter estimates, and the probability integral transform, the registered data are transformed to standard Laplace scale, independently for each quantity at each registration location. The resulting Laplace-scale data are shown in Figure~\ref{Fgr4}, and form the input for the MSCE inference discussed in Sections~\ref{Sct:Mth} and ~\ref{Sct:Rsl} below. The characteristics of Figure~\ref{Fgr4} are discussed further in motivating the results in Sections~\ref{Sct:Rsl}.
\begin{figure}
	\centering
	\includegraphics[width=0.8\columnwidth]{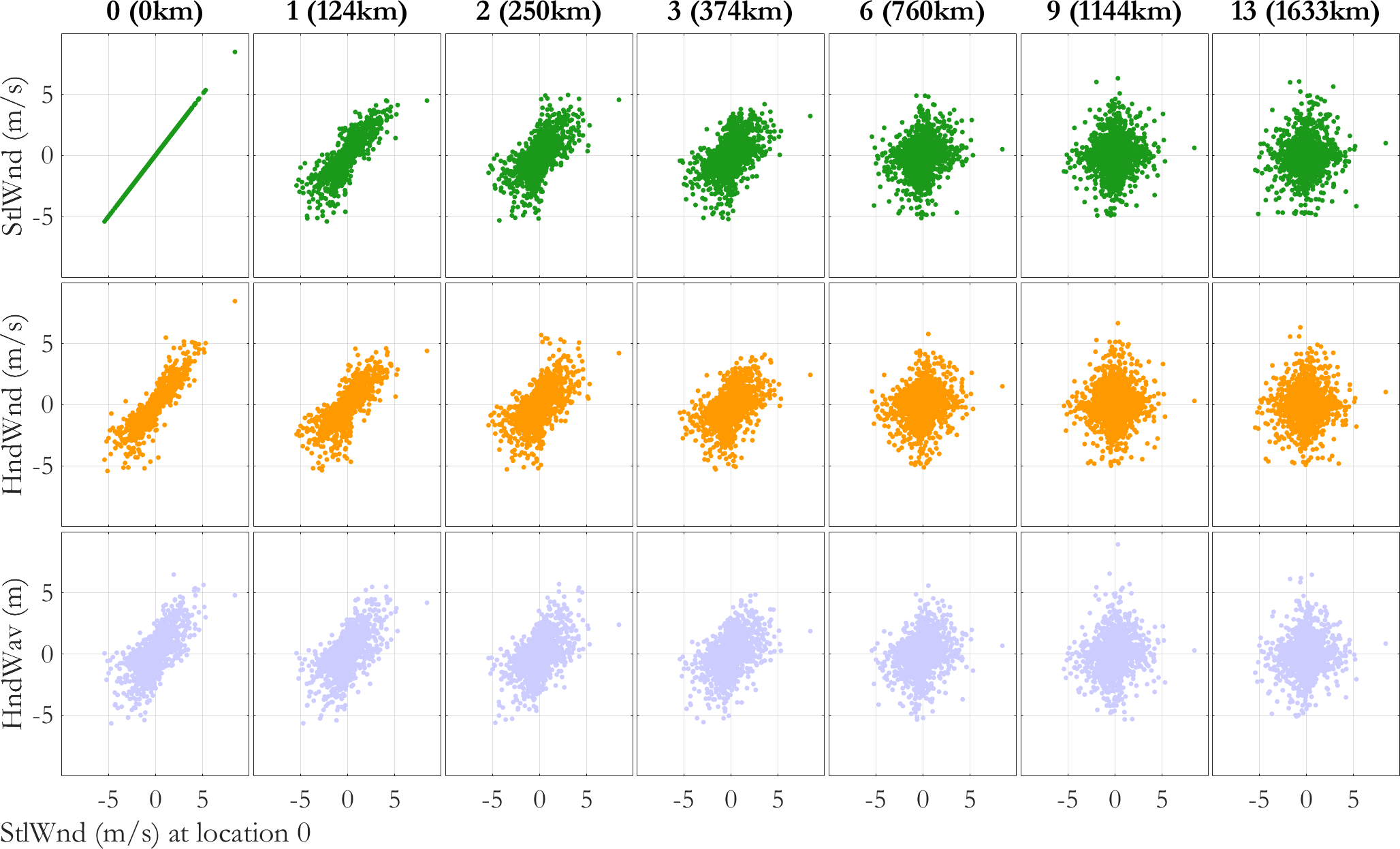}		
	\caption{Scatter plots of \texttt{StlWnd} (green), \texttt{HndWnd} (orange) and \texttt{HndWav} (blue; all on standard Laplace scale) at selected locations $r_j$, $j=0, 1, 2, 3, 6, 9, 13$ against \texttt{StlWnd} at the conditioning location $r_j$, $j=0$. See Figure~\ref{Fgr2} for the corresponding plot on physical scale.}
	\label{Fgr4}
\end{figure}

The conditional extremes model of \cite{HffTwn04} and derivatives is motivated by asymptotic arguments. Such arguments in multivariate extreme value theory generally assume that variables have common marginal distributions. The conditional extremes model form is most simply represented when that marginal distribution is the standard Laplace (\citealt{KefPpsTan13}). From an applications perspective, the marginal transformation from physical to Laplace scales, achieved using the PPC model, also allows the effects of covariates to be captured. Hence, on Laplace scale following transformation using a non-stationary marginal extreme value model, the data are likely to exhibit little or no covariate dependence, and inference is likely to me more straightforward.

\section{Methodology} \label{Sct:Mth}

\subsection{Introduction} \label{Sct:Mth:Int}
The multivariate spatial conditional extremes model is an extension of a sequence of earlier models for conditional extremes (\citealt{HffTwn04}) and spatial conditional extremes (\citealt{TwnEA18}, \citealt{ShtEA19}, \citealt{WdsTwn19}, \citealt{ShtEA20}, \citealt{ShtEA20a}). 

The underlying idea for all these models is a description of the conditional distribution of one or more conditioned variates $Y$ given a large value $x>u$ of a conditioning variate $X$ from \cite{HffTwn04}, motivated by asymptotic arguments as $u \rightarrow \infty$. For variables expressed on standard Laplace scale with positive association (as explained in \citealt{KefTwnLmb13}), this relationship can be written
\begin{eqnarray*}
Y|\{X=x\}=\alpha x + x^\beta Z, \quad x>u
\end{eqnarray*}
for $\alpha \in (0,1]$, $\beta \in (-\infty,1]$, where the residual $Z$ follows some unknown non-degenerate distribution $G$. For the purposes of parameter estimation, it is commonly assumed that $Z \sim N(\mu, \sigma^2)$ for $\mu \in \mathbb{R}$ and $\sigma>0$, a normal distribution. Inference therefore involves estimation of $\alpha$, $\beta$, $\mu$ and $\sigma$.

The methodology can be extended to multivariate conditional extremes by considering a vector $\un{Y}=(Y_1,Y_2,...,Y_m)$ of $m$ conditioned variates, with the model form becoming $\un{Y}|\{X=x\}=\un{\alpha} x + x^{\un{\beta}} \un{Z}$ for $x>u$. Vectors $\un{\alpha}$, $\un{\beta}$ now have elements $\alpha_k \in (0,1]$, $\beta_k \in (-\infty,1]$, and $\un{Z} \sim G$ now represents the joint residual over $m$ conditioned quantities. We might approximate this using a multivariate normal distribution $N(\un{\mu}, \un{\Sigma})$ with mean vector $\un{\mu}$ and covariance matrix $\un{\Sigma}$. Inference then involves estimation of $\un{\alpha}$, $\un{\beta}$, $\un{\mu}$ and $\un{\Sigma}$.

The methodology can also be extended to a spatial context, as described by \cite{ShtEA19} and \cite{WdsTwn19}, by considering a stationary spatial process $X(\cdot)$ on domain $\mathbb{R}$ with standard Laplace marginal distributions. Then for a conditioning location $r_0$ and remote location $r$ with separation $d=|r-r_0|$, we might write $X(r) \vert \{X(r_0)=x\}=\alpha(d) x+x^{\beta(d)}Z(d)$ for $x>u$. Inference for a finite set of locations $r_0, r_1, ..., r_p$ reduces to multivariate conditional extremes with a particular choice of parametric form for residual process $\un{Z}$, to encode the fact that as $d$ increases, the dependence between locations will typically decrease to zero in environmental applications. In particular, for locations $r_0, r$ corresponding to large $d$, we expect that $\alpha(d) \rightarrow 0$ and $\beta(d) \rightarrow 0$ so that $Z(d)$ must follow the standard Laplace distribution marginally. However, for small $d$, a multivariate normal choice for $G$ might still be more appropriate. For this reason, a choice for $G$ with marginal delta-Laplace (or generalised Gaussian) distribution is useful, admitting marginal standard normal and standard Laplace forms \citep{WdsTwn19}. The residual dependence structure of $G$ might be represented by a (conditional) Gaussian field, as explained further in Section~\ref{Sct:Mth:Otl}.

\subsection{Outline of model} \label{Sct:Mth:Otl}
The MSCE model incorporates aspects of both multivariate conditional extremes and spatial conditional extremes. From the perspective of inference for observations at a finite set of locations $r_0, r_1, ..., r_p$, the MSCE model reduces to the multivariate conditional extremes model with a specific parametric choice for the distribution $G$ of the residual process $\un{Z}$.

Suppose that random variable $X_{j,k}$ represents quantity $k=1,2,...,m$ at location $r_j$, $j=0,1,...,p$ on standard Laplace scale, and that $\un{X}$ represents the set of ``remote'' random variables $X_{1,1}, X_{2,1}, ..., X_{p,1}$, $X_{1,2}, X_{2,2}, ..., X_{p,2}$, $..., X_{1,m}, ..., X_{p,m}$. For brevity, we also write $\un{X}$ as $\{X_{j,k}\}$ for $(j,k) \in \mathcal{I}_{\text{Rmt}}$, for the ordered set 
\begin{eqnarray*}
\mathcal{I}_{\text{Rmt}} = \{(1,1), (2,1),...(p,1),(1,2),...,(p,2), ..., (1,m),...,(p,m)\}
\end{eqnarray*}
and define the function $\mathcal{A}(j,k)=p(k-1)+j$ which returns the location of the pair $(j,k)$ in the ordered set $\mathcal{I}_{\text{Rmt}}$. Then, for a large value $x$ of the conditioning random variable $X_{0,1}$ corresponding to quantity $k=1$ at location $r_0$, we assume we can write
\pben
\un{X} | \{X_{0,1}=x\} = \un{\alpha} x + x^{\un{\beta}} \un{Z}, \quad x>u
\label{E:BasicMSCE}
\peen
where the vector of parameters $\un{\alpha}$ has elements $\alpha_{\mathcal{A}(j,k)} \in  (0,1]$ for $(j,k) \in \mathcal{I}_{\text{Rmt}}$, the exponent vector $\un{\beta}$ has elements $\beta_{\mathcal{A}(j,k)} \in (-\infty,1]$, and element-wise multiplication of terms is assumed. We assume further that the residual process $\un{Z}$ follows a delta-Laplace distribution with conditional Gaussian covariance structure, parameterised as $\un{Z} \sim \text{DL}(\un{\mu},\un{\sigma}^2,\un{\delta};\un{\Sigma})$ for mean vector $\un{\mu}$ with elements $\mu_{\mathcal{A}(j,k)} \in \mathbb{R}$, marginal variance vector $\un{\sigma}^2$ with elements $\sigma_{\mathcal{A}(j,k)}^2>0$, delta-Laplace parameter vector $\un{\delta}$ with elements $\delta_{\mathcal{A}(j,k)}>0$. The marginal density function for any $(j,k) \in \mathcal{I}_{\text{Rmt}}$ is given by 
\begin{eqnarray}
f_{Z_{j,k}}(z_{j,k})=\frac{\delta_{j,k}}{2\kappa_{j,k}\sigma_{j,k}\Gamma\left(\frac{1}{\delta_{j,k}}\right)}\exp\left\{-\left|\frac{z-\mu_{j,k}}{\kappa_{j,k}\sigma_{j,k}}\right|^{\delta_{j,k}}\right\}
\label{E:ZDns}
\end{eqnarray}
where $\kappa_{j,k}^2=\Gamma\left(1/\delta_{j,k}\right)/\Gamma\left(3/\delta_{j,k}\right)$ and $\Gamma(\cdot)$ represents the gamma function. The mean and variance of this distribution are respectively $\mu_{j,k}$ and $\sigma_{j,k}^2$, regardless of the choice of $\delta_{j,k}$. The dependence structure of $\un{Z}$ is described on Gaussian scale, with correlation matrix
\pbe
\un{\Sigma}_{\mathcal{A}(j,k)\mathcal{A}(j',k')} =  \frac{\un{\Sigma}_{\mathcal{A}^*(j,k)\mathcal{A}^*(j',k')}^* - \un{\Sigma}_{\mathcal{A}^*(j,k)\mathcal{A}^*(0,1)}^* \un{\Sigma}_{\mathcal{A}^*(0,1)\mathcal{A}^*(j',k')}^*} {\left(1-\un{\Sigma}_{\mathcal{A}^*(j,k)\mathcal{A}^*(0,1)}^{*2}\right)^{1/2}\left(1-\un{\Sigma}_{\mathcal{A}^*(0,1)\mathcal{A}^*(j',k')}^{*2}\right)^{1/2}}
\pee
for $(j,k)$ and $(j',k')$ $\in \mathcal{I}_{\text{Rmt}}$. The correlation matrix is simply that of a standard Gaussian field evaluated at the $p$ locations $r_j$. $j=1,...,p$ for all $m$ quantities, conditioned on the value of one quantity ($k=1$) at an external location $r_0$. The correlation matrix for the corresponding unconditioned standard Gaussian field is
\pben
\un{\Sigma}_{\mathcal{A}^*(j,k)\mathcal{A}^*(j',k')}^* = \lambda_{k,k'}^{|k-k'|} \exp\left( - \left(\frac{\text{dist}(r_{j},r_{j'})}{\rho_{k,k'}}\right)^{\kappa_{k,k'}}\right)
\label{E:UncGssFld}
\peen
for pairs $(j,k)$ and $(j',k')$ in the extended ordered set $\mathcal{I}^*_{\text{Rmt}} = \mathcal{I}_{\text{Rmt}}\cup \{(0,1)\}$ (i.e. including the conditioning quantity $k=1$ at conditioning location $r_0$), where function $\mathcal{A}^*(j,k)$ now identifies the location of pair $(j,k)$ in $\mathcal{I}^*_{\text{Rmt}}$. Parameters $\lambda_{k,k'} \in [0,1]$ are the assumed common correlations between quantities at any one location for $k \neq k'$, and $\lambda_{k,k'}=1$ when $k=k'$. Further $\rho_{k,k'}>0$ and $\kappa_{k,k'}>0$ are the scale and exponent parameters of the assumed powered exponential dependence of the standard Gaussian field for quantities $k$ and $k'$. We choose to write the unique set of parameters $\{\lambda_{k,k'}\}_{k'>k}$ for estimation as vector $\un{\lambda}$, and the unique sets  $\{\rho_{k,k'}\}_{k' \ge k}$ and $\{\kappa_{k,k'}\}_{k' \ge k}$ as vectors $\un{\rho}$ and $\un{\kappa}$ respectively. $\text{dist}(\un{r}_{j},\un{r}_{j'})$ is the distance between locations $\un{r}_j$ and $\un{r_{j'}}$, calculated in metres using spherical distance on the Earth's surface.

Thus the joint distribution $G$ of $\un{Z}$ can be written
\begin{eqnarray*}
    G(\un{z})=\Phi_{mp}\left(\left(\Phi^{-1}(F_{Z_{1,1}}(z_{1,1})),\Phi^{-1}(F_{Z_{2,1}}(z_{2,1})),\ldots,\Phi^{-1}(F_{Z_{j,k}}(z_{j,k})),\ldots,\Phi^{-1}(F_{Z_{p,m}}(z_{p,m}))\right);\mathbf{0},\boldsymbol{\Sigma}\right)
\end{eqnarray*}
where $\Phi$ is the cumulative distribution function of a standard Gaussian distribution, and $\Phi_{mp}(\un{z};\mathbf{0},\mathbf{\Sigma})$ is the cumulative distribution function of a $mp$-dimensional Gaussian distribution with mean $\mathbf{0}$ and covariance matrix $\mathbf{\Sigma}$ evaluated at $\un{z}$. $F_{j,k}(z)$ is the delta-Laplace marginal cumulative distribution function for quantity $k$ at location $j$ with density given by Equation~\ref{E:ZDns}. By differentiating the expression for $G(\un{z})$, an expression for the log-density of $\un{Z}$ can be found, and hence an expression for the sample log-likelihood required for inference (see \citealt{ShtEA20}).

\subsection{Inference} \label{Sct:Mth:Inf}
Inference involves estimation of parameter vectors $\un{\alpha}$, $\un{\beta}$, $\un{\mu}$, $\un{\sigma}$, $\un{\delta}$, and $\un{\lambda}$, $\un{\rho}$ and $\un{\kappa}$. The variation of any one of $\alpha$, $\beta$, $\mu$, $\sigma$ and $\un{\delta}$ with distance is described using a piecewise linear representation (with parameters estimated at each of $n_\text{Nod}$ nodes) for each of the $m$ quantities of interest. Thus for generic parameter $\eta$ (i.e. any of ${\alpha}$, ${\beta}$, ${\mu}$, ${\sigma}$, ${\delta}$ for a specific quantity, varying with distance $d$), with node values $\{\eta_\ell^N\}_{\ell=1}^{n_\text{Nod}}$, the assumed piecewise linear representation is $\eta(d)=(h_L \eta_{\ell^*}^N + h_U \eta_{\ell^*+1}^N)/h$, where $h_L=d-d_{k^*}$, $h_U=d_{k^*+1}-d$ and $h=d_{k^*+1}-d_{k^*}$, and $k^*=\argmax{k}(d_k:d_k<d)$. The full inference therefore requires the estimation of parameter set $\Omega$ given by
\begin{eqnarray*}
\Omega=\left(\{\alpha^N_{\ell,k},\beta^N_{\ell,k},\mu^N_{\ell,k},\sigma^N_{\ell,k},\delta^N_{\ell,k}\}, \{\lambda_{k,k'}\}_{k'>k}, \{\rho_{k,k'},\kappa_{k,k'}\}_{k' \ge k} \right), \quad \ell=1,2,..., n_{\text{Nod}}, \quad k,k'=1,2,...,m.
\end{eqnarray*}
Noting that the dimensions of $\un{\lambda}$, $\un{\rho}$ and $\un{\kappa}$ are respectively $m(m-1)/2$, $m(m+1)/2$ and $m(m+1)/2$, $\Omega$ therefore contains $m(5 n_\text{Nod}+(3m+1)/2)$ parameters.

We use Bayesian inference to estimate the joint posterior distribution of MSCE model parameters. An adaptive MCMC algorithm based on \cite{RbrRsn09} is used for parameter inference, described in \cite{ShtEA19}, \cite{ShtEA20} and \cite{ShtEA20a}. Briefly, random search is used to find a reasonable starting solution. Then a Metropolis-within-Gibbs algorithm is used iteratively to sample each of the individual parameters in turn for a total of $n_1=250$ iterations. Subsequently we use the adaptive MCMC algorithm to update all parameters jointly for a further $n_2=19750$ iterations. 

Uniform prior distributions were assumed for each parameter; it was confirmed that posterior densities were not obviously restricted by prior specification. In particular, we allow the node values of $\alpha$ to be $>1$ (as discussed by \citealt{TndEA21} for sub-asymptotic levels), and hence did not adopt the conditional quantile constraints of \cite{KefPpsTan13}. 

Data and prototype MATLAB code for the analysis discussed in this article are available at \cite{ShtRssJnt21}.

\section{Results} \label{Sct:Rsl}
Results of applying the MSCE model to the Laplace-scale sample illustrated in Figure~\ref{Fgr4} are now discussed. Inspection of Figure~\ref{Fgr4} provides some guidance regarding the decay of $\alpha$ with distance we might expect to infer. Consider the scatter plot of \texttt{StlWnd} at location $r_1$ on \texttt{StlWnd} at conditioning location $r_0$ in the first row and second column of the figure. For large values (say $\ge 4$) of \texttt{StlWnd} at $r_0$, the values of \texttt{StlWnd} at $r_1$ are also relatively large, suggesting that the corresponding value of $\alpha$ estimated (see Equation~\ref{E:BasicMSCE}) should be near unity. As we move across the first row to larger distances $r_j$, it is clear that typical values of \texttt{StlWnd} at $r_j$ (for large values of \texttt{StlWnd} at $r_0$) are centred around zero, suggesting that $\alpha$ at these locations will be near zero. Indeed, for all of \texttt{StlWnd}, \texttt{HndWnd} and \texttt{HndWav}, it appears that the value of $\alpha$ for all quantities will be near zero for locations $r_j$ with $j \ge 6$, or $r_j>760$ km. At a smaller distance $r_j$ (with $j \le 3$), the joint characteristics of the three quantities of interest appear rather similar on Laplace scale. 

\begin{figure}
	\centering
	\includegraphics[width=0.8\columnwidth]{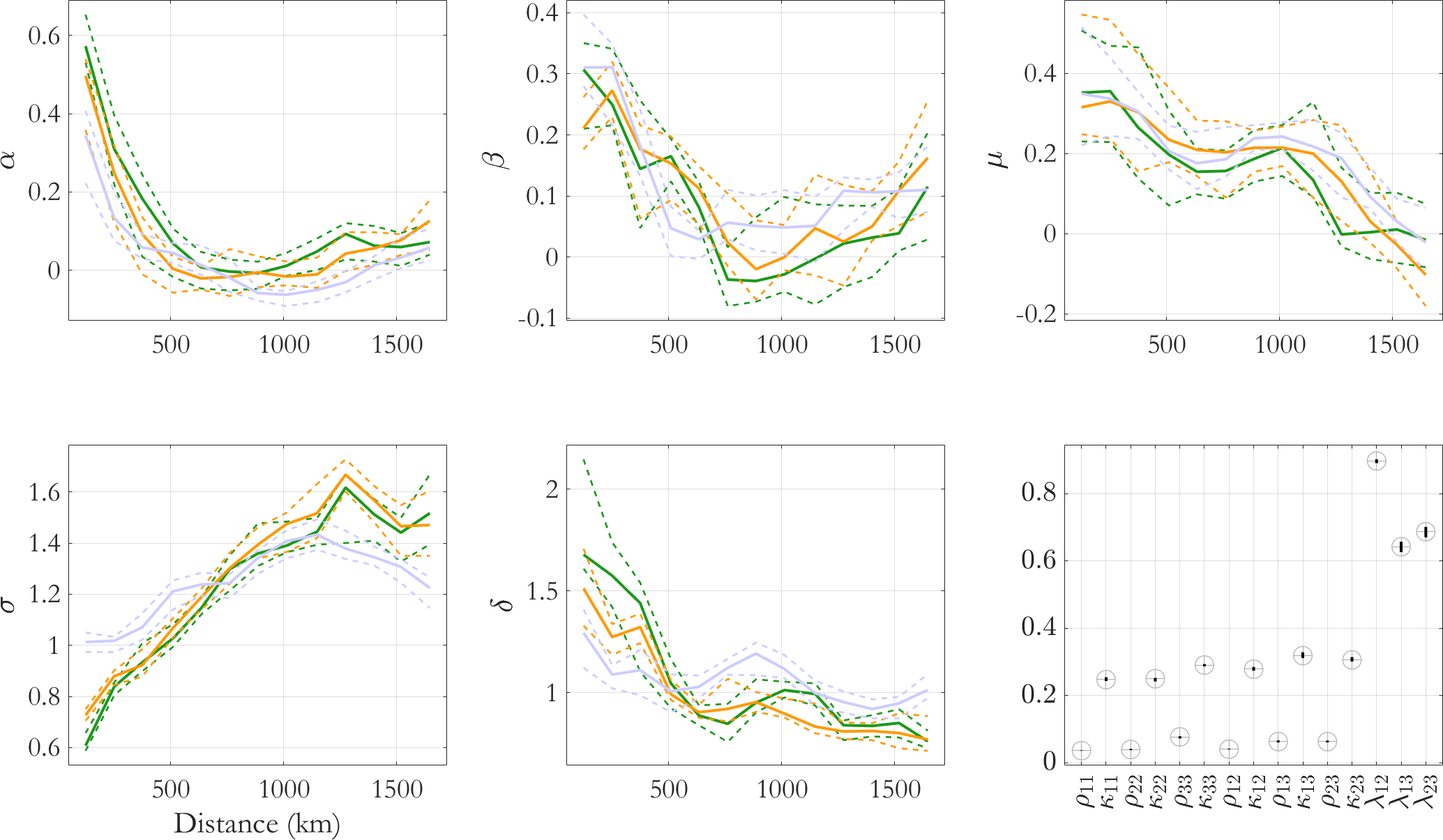}		
	\caption{Posterior parameter estimates (shown as mean (solid lines) and 95\% credible intervals (dashed lines)) for marginal MSCE model parameters $\alpha$, $\beta$, $\mu$, $\sigma$, $\delta$, and residual dependence parameters $\rho$, $\kappa$ and $\lambda$. Conditioning on \text{StlWnd} at location $r_0$ with conditioning value equal to the 0.75 quantile of the standard Laplace distribution. Colour coding indicates the conditioned quantity: \texttt{StlWnd} (green), \texttt{HndWnd} (orange), and \texttt{HndWav} (blue). Estimates for residual dependence parameters are given by large circles, centred on the posterior mean estimate, and vertical black lines reflecting posterior 95\% credible intervals. For further interpretation of residual dependence parameters, see Equation~\ref{E:UncGssFld}.}
	\label{Fgr5}
\end{figure}

The corresponding posterior estimates for $\alpha$, together with those for $\beta$, $\mu$, $\sigma$ and $\delta$ are shown in Figure~\ref{Fgr5}. As anticipated $\alpha$ decays from a value $>0$ at $r_1$ to around zero for distances over 600 km. Estimates for $\alpha$ are generally somewhat larger for \texttt{StlWnd} than for \texttt{HndWnd} than for \texttt{HndWav} as might be expected, but the differences are small given the widths of credible intervals. It is interesting that the profiles for $\mu$ with distance are also similar for the three quantities, reducing from around 0.4 for small distances to around zero for large distances. Estimates for $\beta$ reduce from approximately 0.3 at small distances to zero or 0.1 with increasing distance. Estimates for $\sigma$ increase towards approximately $\sqrt 2$ with increasing distance, as suggested by the model formulation: for large distances, the effect of conditioning on $X_{0,1}$ is negligible, so that $X_{j,k}|\{X_{0,1}=x\}$ is similar to the unconditioned $X_{j,k}$ which is standard Laplace distributed, with variance equal to 2. The behaviour of $\sigma$ for \texttt{HndWav} is somewhat different, reflecting the difference between $H_S$ and the other two wind speed variates. The decay of $\delta$ with increasing distance to around unity, suggests that $X_{j,k}|\{X_{0,1}=x\}$ is Laplace-distributed (with $\delta=1$) for large distances, but more Gaussian (with $\delta=2$) for small distances, as expected.

Parameter estimates for the residual dependence structure (see Equation~\ref{E:UncGssFld}) in the bottom right panel indicate rather similar shape and scale estimates for distance decay of pairs of residuals for \texttt{StlWnd} ($\rho_{11}, \kappa_{11}$), pairs of residuals for \texttt{HndWnd} ($\rho_{22}, \kappa_{22}$), and pairs of residuals for \texttt{HndWav} ($\rho_{33}, \kappa_{33}$). Shape and scale parameters for cross-dependence between different quantities take similar values, although it is noteworthy that residual parameters involving \texttt{HndWav} tend to be somewhat larger. We note that the values of $\rho$ and $\kappa$ have been scaled so that they fall comfortably in the interval [0,1] for convenience during MCMC inference. The actual values of these parameters (e.g. appropriate for input to Equation~\ref{E:UncGssFld}) are given by $100\rho$ and $5\kappa$ respectively. The large value for $\lambda_{1,2} \approx 0.9$ indicates high zero-distance residual correlation between \texttt{StlWnd} and \texttt{HndWnd}. Similar smaller values for $\lambda_{1,3}$ and $\lambda_{2,3}$ around 0.7 indicate lower zero-distance residual correlation between a wind speed variate and the $H_S$ variate \texttt{HndWnd}, again as might be expected from physical considerations.

\begin{figure}
	\centering
	\includegraphics[width=0.8\columnwidth]{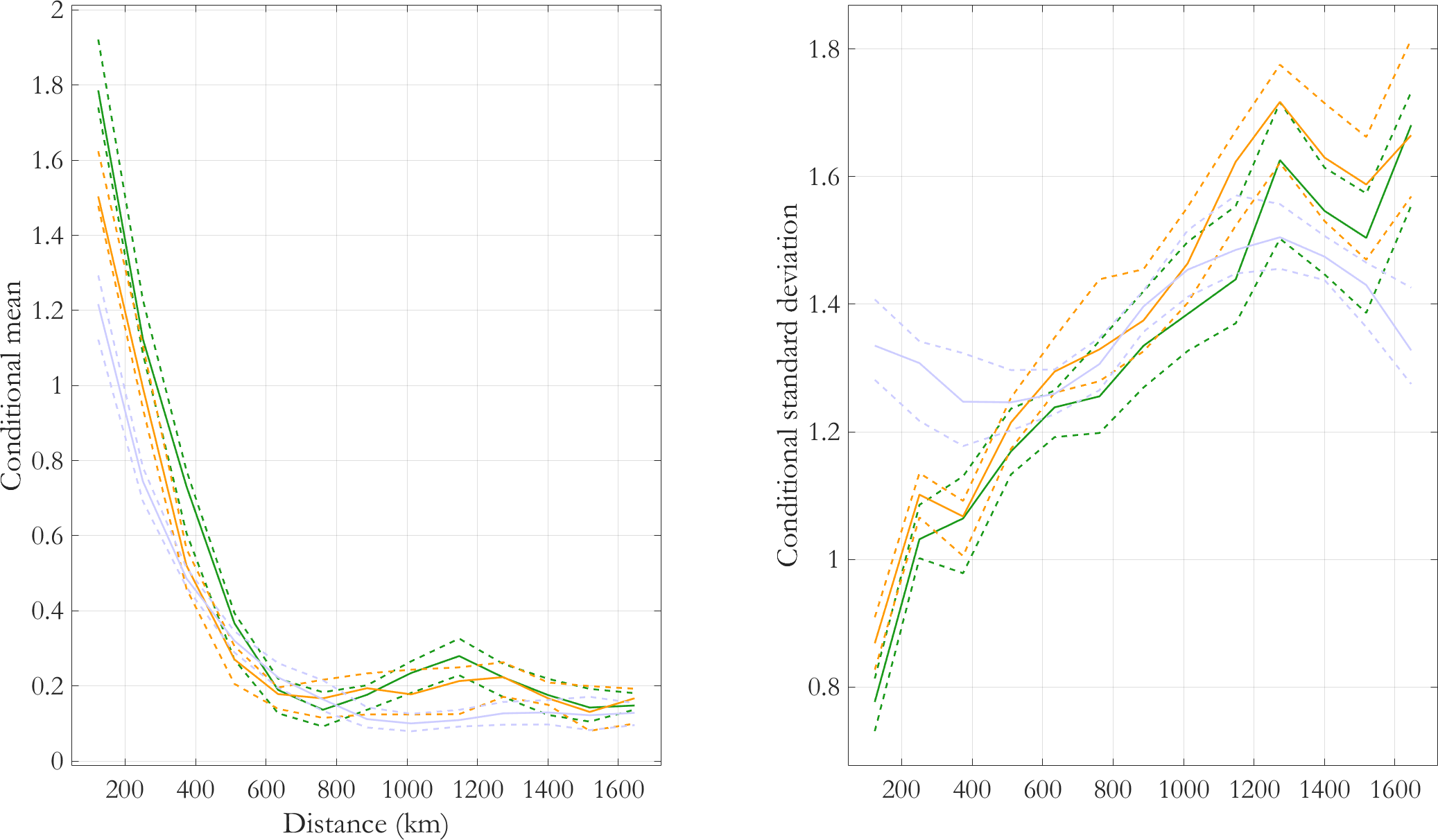}		
	\caption{Conditional mean ($\alpha x + x^\beta \mu$) and standard deviation ($\sigma x^\beta$, both shown as mean and 95\% credible intervals) from estimated MSCE model for conditioning on \text{StlWnd} at location $r_0$ with conditioning value equal to the 0.95 quantile of the standard Laplace distribution. Colour coding indicates the conditioned quantity: \texttt{StlWnd} (green), \texttt{HndWnd} (orange), and \texttt{HndWav} (blue).}
	\label{Fgr6}
\end{figure}

Figure~\ref{Fgr6} shows conditional mean profiles for \texttt{StlWnd} (green), \texttt{HndWnd} (orange) and \texttt{HndWav} (blue) estimated under the fitted MSCE model, for conditioning on $X_{0,1}$ with conditioning value corresponding to the 0.95 quantile of the standard Laplace distribution, at approximately 2.3. The characteristics of the conditional mean are rather similar to those of $\alpha$ and $\mu$ in Figure~\ref{Fgr5}. At small distances, the conditional mean of \texttt{StlWnd} is clearly larger than that for \texttt{HndWnd}, which is itself larger than for \texttt{HndWav}. However, after approximately 600 km, all conditional mean profiles have decayed to a value of around 0.2. The profiles of conditional standard deviation with distance reflect the characteristics of $\sigma$ in Figure~\ref{Fgr5}.

\begin{figure}
	\centering
	\includegraphics[width=0.8\columnwidth]{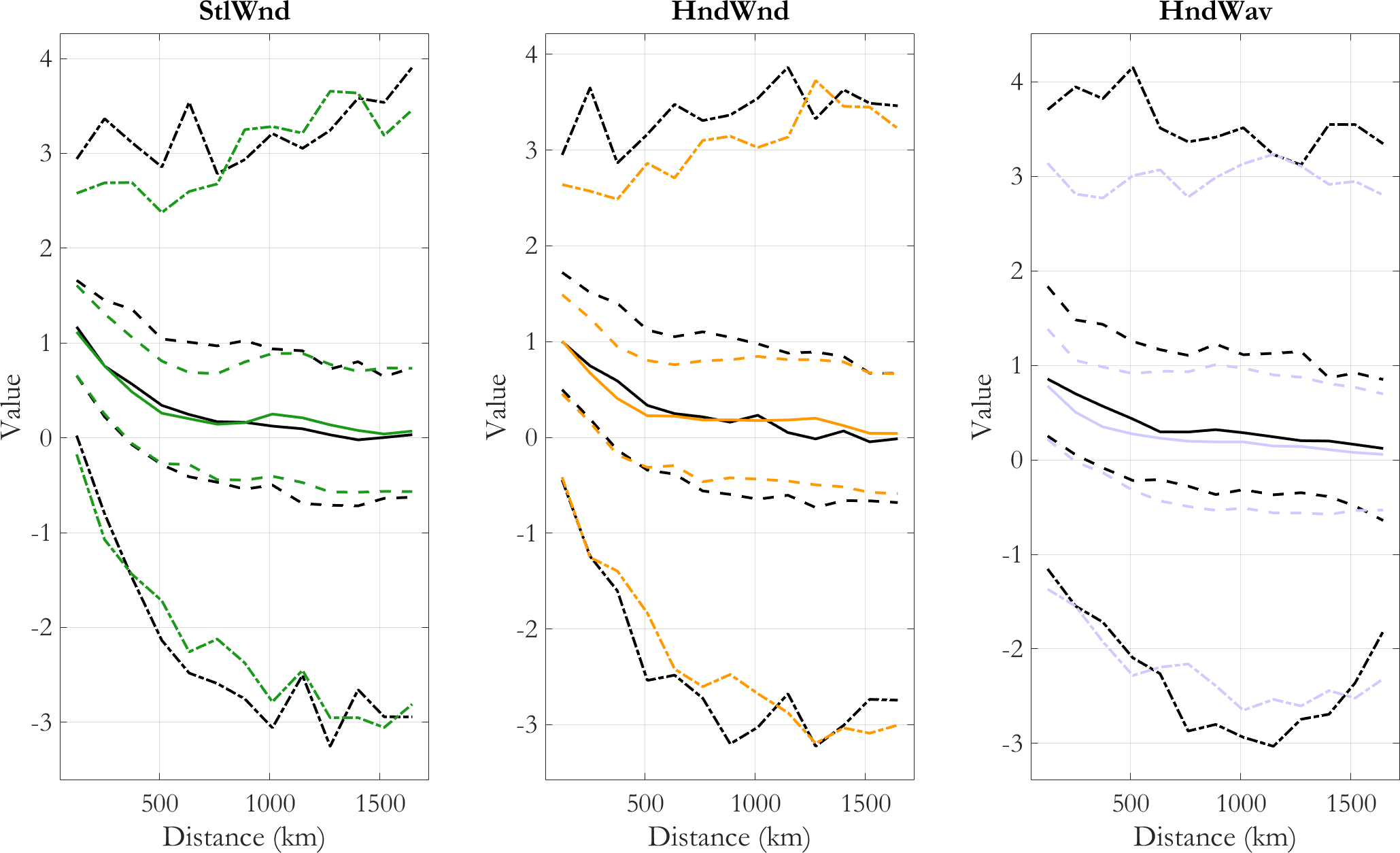}		
	\caption{Model validation. Comparison of observed (Laplace-scale) data with distance with simulations under the fitted MSCE model. Conditioning on \text{StlWnd} at location $r_0$ with conditioning value equal to the 0.75 quantile of the standard Laplace distribution. Shown are 0.025, 0.25, 0.5, 0.75 and 0.975 quantiles of the observations (black) and simulation. Colour coding indicates the conditioned quantity for simulation: \texttt{StlWnd} (green), \texttt{HndWnd} (orange), and \texttt{HndWav} (blue).}
	\label{Fgr7}
\end{figure}

\begin{figure}
	\centering
	\includegraphics[width=0.8\columnwidth]{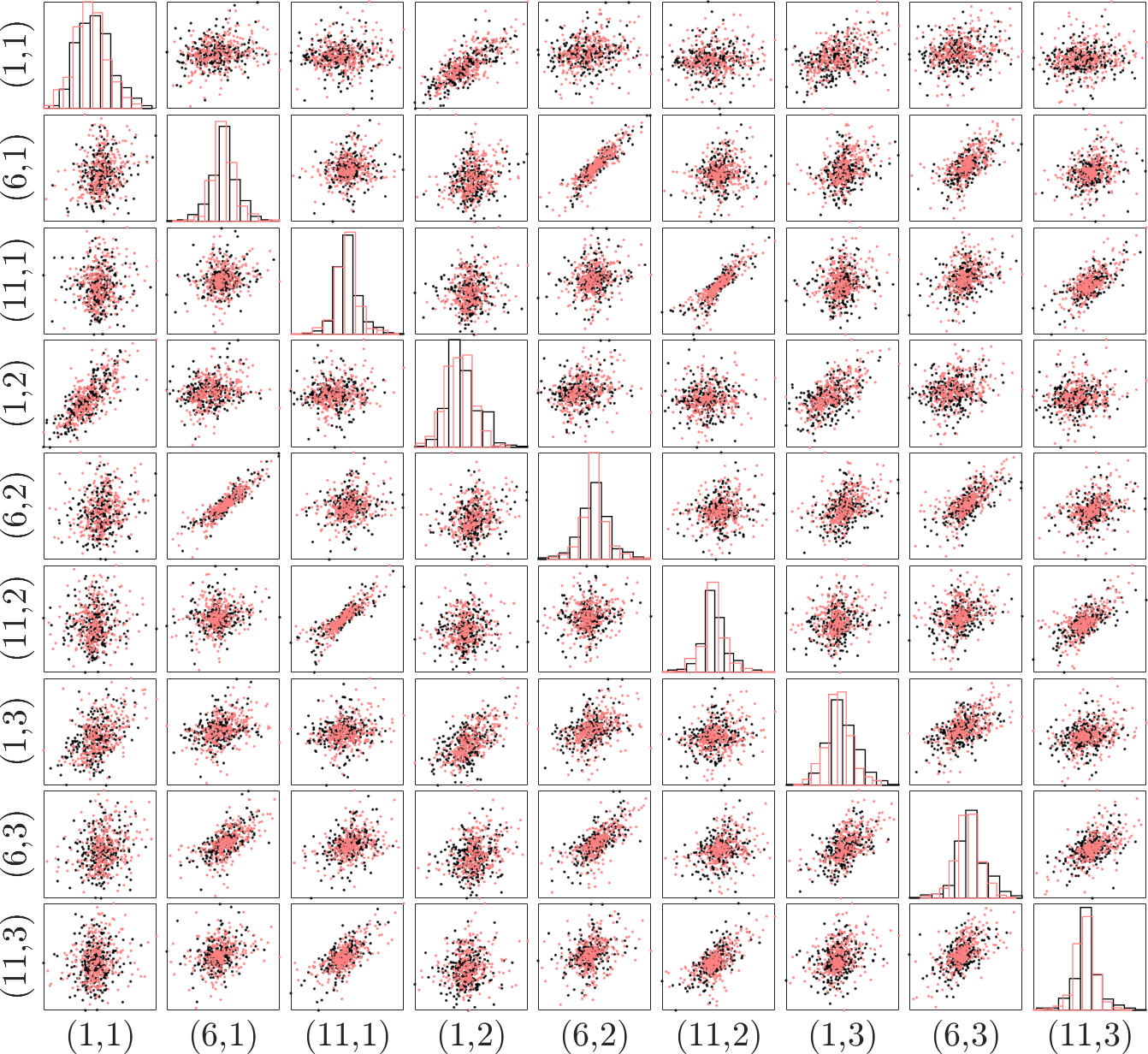}		
	\caption{Model validation. Comparison of observed residuals (black) and residuals simulated under the fitted MSCE model (red). Diagonal panels show histograms of residuals for selected location-quantity pairs $(j,k) \in \mathcal{I}_{\text{Rmt}}$. Off-diagonal panels show scatter plots of residuals for selected pairs $(j,k), (j',k') \in \mathcal{I}_{\text{Rmt}}$. Conditioning on \text{StlWnd} at location $r_0$ with conditioning value equal to the 0.75 quantile of the standard Laplace distribution.}
	\label{Fgr8}
\end{figure}

Figures~\ref{Fgr7} and \ref{Fgr8} explore the quality of fit of the MSCE model to the sample. Figure~\ref{Fgr7} shows 0.025, 0.25, 0.5, 0.75 and 0.975 quantiles from marginal simulations for \texttt{StlWnd} (green), \texttt{HndWnd} (orange) and \texttt{HndWav} (blue) with distance for conditioning on \text{StlWnd} at location $r_0$ with conditioning value equal to the 0.75 quantile of the standard Laplace distribution. Also shown in black are the corresponding quantiles calculated directly from the sample. There is good agreement, indicating that marginally at least the model is able to capture the distance-dependent features of the data.

Figure~\ref{Fgr8} compares the characteristics of observed residuals (black) and corresponding residuals simulated under the fitted model (red). Diagonal panels compare histograms for selected (location,quantity) pairs, and off-diagonal panels show scatter plots for different (location,quantity) pairs. Conditioning is again on \text{StlWnd} at location $r_0$ with conditioning value equal to the 0.75 quantile of the standard Laplace distribution. The quality of agreement between the empirical distributions of observed and simulated residuals was also quantified using bootstrapping to estimate a null distribution for the Kullback-Leibler divergence between the distributions of bootstrap resamples of the observed residuals. Then the tail probability corresponding to the Kullback-Leibler divergence between the distributions of observed and simulated residuals in the null distribution is estimated. We found that approximately 20\% of the values exceeded the 95\% percentile of the null distribution. This indicates that there is reasonable if not excellent agreement between the samples of observed and simulated residuals, and that the residual dependence model is able to capture the sample characteristics. 

\section{Discussion} \label{Sct:Dsc}
This article outlines a multivariate spatial conditional extremes (MSCE) model to describe the dependence between extremes of multiple different spatial fields. The model is the latest extension of the conditional extremes model of \cite{HffTwn04}. The MSCE model quantifies the conditional distribution of a number of quantities measured on a common set of locations, conditional on the occurrence of a large value of one of the quantities at a conditioning location. The key characteristic of the model is smooth variation of marginal and residual dependence with increasing distance between spatial locations. Bayesian inference is used for parameter estimation. The model is applied to observations of wind speed from satellite observation (\texttt{StlWnd}) and hindcast (\texttt{HndWnd}), and hindcast significant wave height (\texttt{HndWav}) on a spatial transect lying between the British Isles and Iceland, conditioning on a large value of \texttt{StlWnd} at the most south-western location on the transect.

There is evidence that the dependence between \texttt{StlWnd} at neighbouring locations is greater than that between  \texttt{StlWnd} and \texttt{HndWnd} at the same neighbouring locations, and that these are both greater than the dependence between  \texttt{StlWnd} and \texttt{HndWav} at the neighbouring locations. However, the overall trends of MSCE model parameters with distance between locations are similar for each of \texttt{StlWnd}, \texttt{HndWnd} and \texttt{HndWnd}. The conditional mean for all quantities decays to a baseline level at a distance of approximately 600 km. We note however that figures in the appendix, considering dependence for extremes of each of \texttt{StlWnd}, \texttt{HndWnd} and \texttt{HndWnd} separately, suggest that the extent of spatial dependence for \texttt{HndWnd} (Figure~\ref{FgrA3}) is somewhat larger than for the other quantities (Figure~\ref{FgrA1} and Figure~\ref{FgrA2}). We speculate that a distance of approximately 600-800km is indicative of the spatial extent of coherence for wind systems in the North East Atlantic. It would be interesting to estimate the corresponding distance for other ocean basins. 

The effect of choice of conditioning quantity, conditioning location and conditioning value were examined for a number of cases, although not exhaustively. Results with similar general characteristics to those reported here were obtained.

The original intention for this work was to combine all of (a) satellite scatterometer measurements for wind speed and direction from Metop, (b) satellite altimeter measurements for significant wave height and wave direction from JASON (e.g. \citealt{ShtEA20a}), and (c) corresponding hindcast data for all the variables in (a) and (b) in one MSCE model. However, it quickly became apparent that the number of approximately joint measurements (given space and time) from Metop and JASON available is small, and so far insufficient for joint modelling from scatterometry and altimetry. 

The current analysis uses satellite observations of average wind speed and direction corresponding to a relatively short period of time on the daily satellite swath over the North Atlantic, and corresponding spatially- and temporally-matched data from the NORA10 hindcast. These observations are not guaranteed to be representative of wind speeds and directions at the locations of interest, since the satellites pass over the North Atlantic at approximately the same time each day, a source of measurement bias. Moreover, the temporal extent of a severe storm in the North Atlantic is of the order of days, meaning that multiple satellite observations from the same storm event are possible from the daily pass for each satellite. These observations are therefore likely to be correlated in time, at least to some extent. Generally, it would be preferable to perform extreme value analysis on ``storm peak'' wind speed and direction, which can reasonably be assumed to be temporally independent; however, these data are not available from the daily passes per satellite.

Inference for the MSCE model is straightforward using the adaptive MCMC algorithm of \cite{RbrRsn09}, and convergence of MCMC chains is relatively rapid; in practice, 10000 MCMC iterations is more than sufficient. We believe that the MSCE methodology is an interesting extension to the statistician's and met-ocean engineer's tool kits, providing a practically applicable yet statistically principled approach to quantification of conditional extremes behaviour over multiple spatial fields. 

%%%%%%%%%%%%%%%%%%%%%%%%%%%%%%%%%%%%%%%%%%%%%%%%%%%%%%
\section{Acknowledgement}
We thank David Randell from Shell for assistance with data preparation. I. R. Young acknowledges ongoing financial support from the Integrated Marine Observing System (IMOS) and the Victorian Government through the Department of Environment, Land, Water and Planning, Australia.  Data and prototype MATLAB code for the analysis discussed in this article are available at \cite{ShtRssJnt21}.
%%%%%%%%%%%%%%%%%%%%%%%%%%%%%%%%%%%%%%%%%%%%%%%%%%%%%%
\newpage
\bibliographystyle{plainnat}
\bibliography{Phil}
%\bibliography{C:/Philip/LaTeX/Bibliography/phil}

%%%%%%%%%%%%%%%%%%%%%%%%%%%%%%%%%%%%%%%%%%%%%%%%%%%%%%
\clearpage % ensure all floats are processed
\setcounter{figure}{0}
\renewcommand{\thefigure}{A\arabic{figure}}
\appendix

\section{Scatter plots of original data}
\begin{figure}
	\centering
	\includegraphics[width=1\columnwidth]{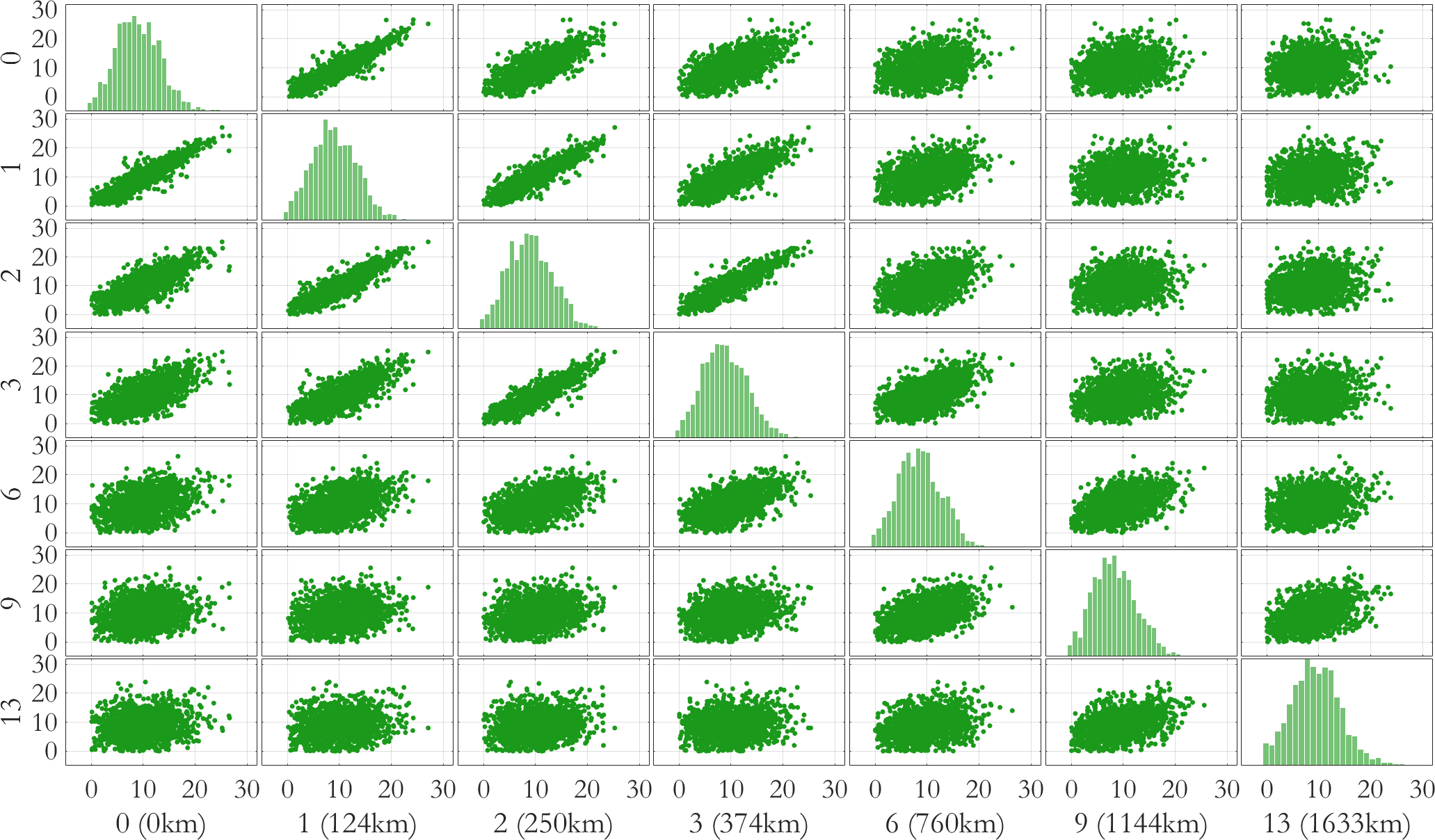}		
	\caption{Off-diagonal entries give scatter plots of \texttt{StlWnd} on physical scale for selected pairs of registration locations $r_j$, $j=0,1,2,3,6,9, 13$ (see Figure\ref{Fgr1}). Diagonal elements give empirical densities for \texttt{StlWnd} at the locations.}
	\label{FgrA1}
\end{figure}

\begin{figure}
	\centering
	\includegraphics[width=1\columnwidth]{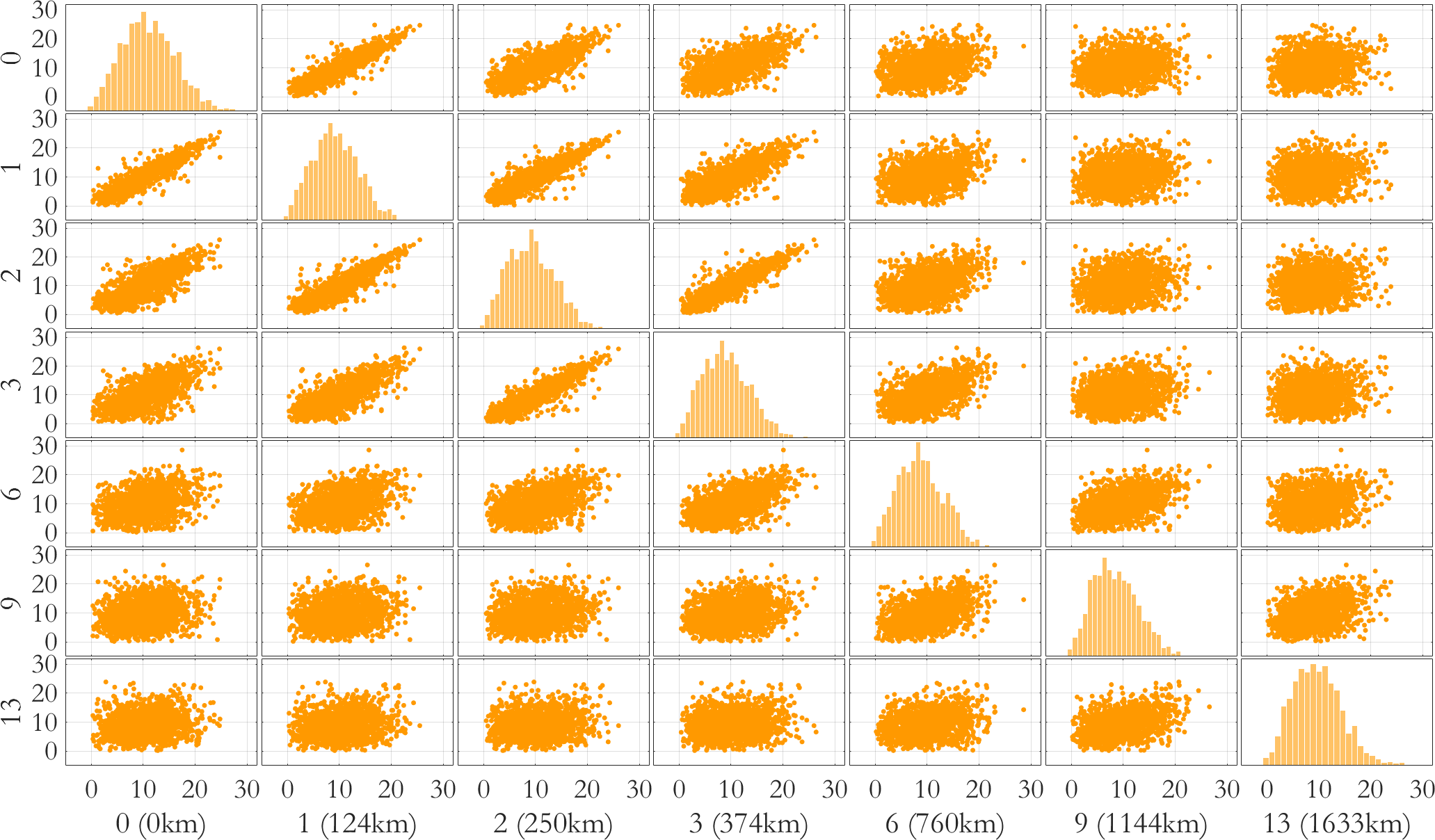}		
	\caption{Off-diagonal entries give scatter plots of \texttt{HndWnd} on physical scale for selected pairs of registration locations $r_j$, $j=0,1,2,3,6,9, 13$. Diagonal elements give empirical densities for \texttt{HndWnd} at the locations.}
	\label{FgrA2}
\end{figure}

\begin{figure}
	\centering
	\includegraphics[width=1\columnwidth]{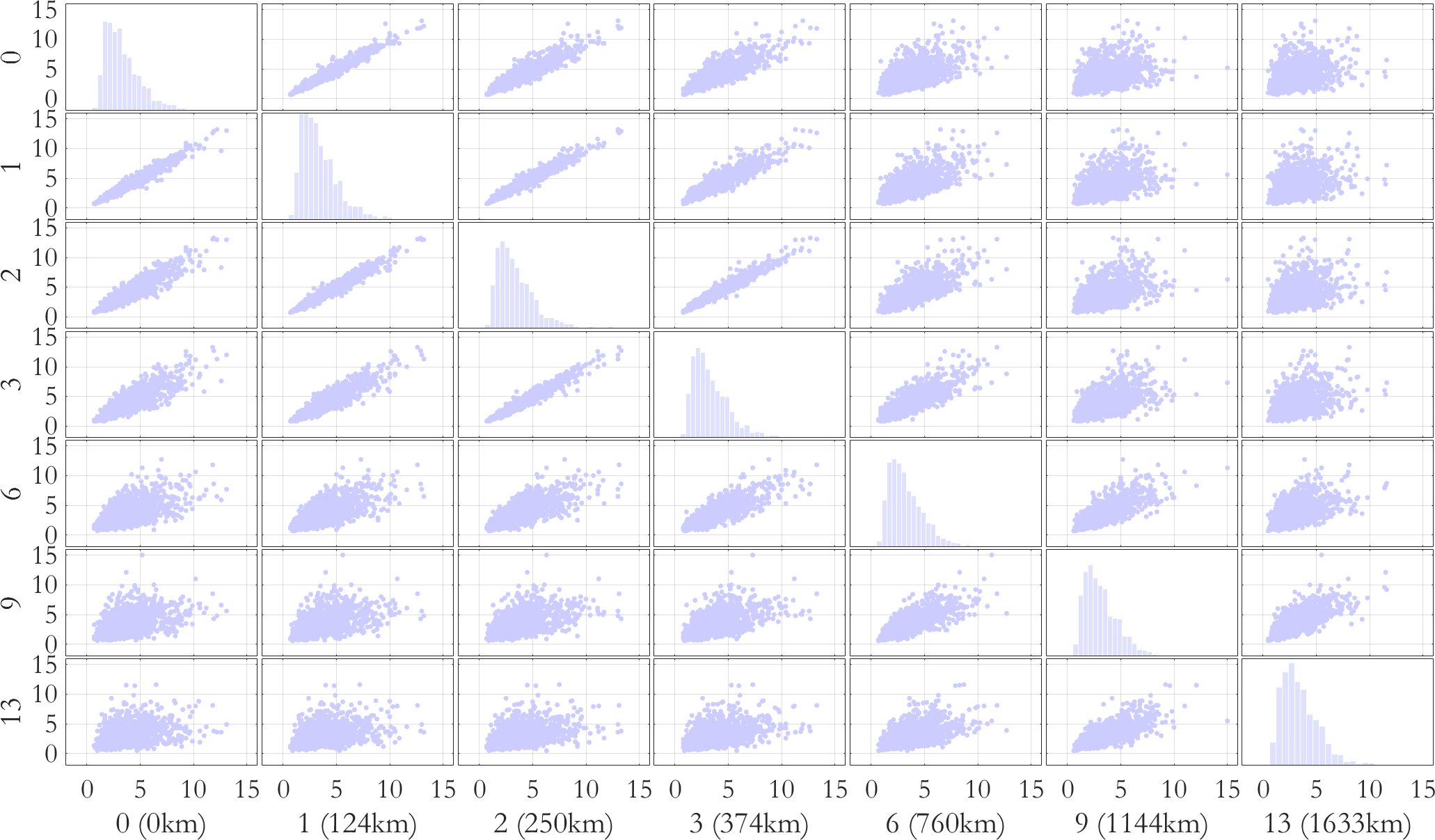}		
	\caption{Off-diagonal entries give scatter plots of \texttt{HndWav} on physical scale for selected pairs of registration locations $r_j$, $j=0,1,2,3,6,9, 13$. Diagonal elements give empirical densities for \texttt{HndWav} at the locations.}
	\label{FgrA3}
\end{figure}

Figures~\ref{FgrA1}-\ref{FgrA3} provide scatter plots and histograms for \texttt{StlWnd}, \texttt{HndWnd} and \texttt{HndWav} on their original physical scales, for selected representative registration locations. The figures reveal a number of interesting features. For example, comparison of Figures~\ref{FgrA1} and \ref{FgrA2} suggests that dependence with distance is rather similar for \texttt{StlWnd} and \texttt{HndWnd}. In contrast, the decay of dependence with distance is more gradual for \texttt{HndWav} as might be expected from physical considerations.

%%%%%%%%%%%%%%%%%%%%%%%%%%%%%%%%%%%%%%%%%%%%%%%%%%%%%%
\end{document}